\documentclass{article}

\usepackage{arxiv}

\usepackage[utf8]{inputenc} 
\usepackage[T1]{fontenc}    
\usepackage{hyperref}       
\usepackage{url}            
\usepackage{booktabs}       
\usepackage{amsfonts}       
\usepackage{nicefrac}       
\usepackage{microtype}      
\usepackage{cite}
\usepackage{lipsum}
\usepackage{setspace}
\usepackage{fancyhdr}       
\usepackage{graphicx}       
\usepackage{pdfpages}
\usepackage{caption}
\usepackage{subcaption}
\graphicspath{{media/}}     

\hypersetup{
    colorlinks=true,       
    linkcolor=blue,        
    citecolor=blue,        
    filecolor=magenta,     
    urlcolor=cyan          
}

\title{Machine-learning designed smart coating: temperature-dependent self-adaptation between a solar absorber and a radiative cooler}

\author{
  Zhaocheng Zhang$^a$, Jiahao Xu$^b$, Pengran Hou$^a$, Yang Deng$^c$\\
  $^a$School of Mechanical Engineering, Dalian University of Technology\\ 
  $^b$School of Mathematics, Dalian University of Technology\\
  $^c$DUT-BSU Joint Institute, Dalian University of Technology\\
  \textit{No.2, Linggong Road, Ganjingzi District, Dalian City, Liaoning Province, China}\\
  Dream020501zzc@mail.dlut.edu.cn \\}

\begin{document}
\maketitle
\onehalfspacing

\begin{abstract}
We designed a multilayered self-adaptive absorber/emitter metamaterial, which can smartly switch between a solar absorber and a radiative cooler based on temperature change. The switching capability is facilitated by the phase change material and the structure is optimized by machine learning. Our design not only advances the machine-learning-based development of metamaterials but also has the potential to significantly reduce carbon emissions and contribute to the goal of achieving carbon neutrality. 
\end{abstract}


\section{Introduction}
Passive temperature adjustment has gained significant attention in recent years due to its potential to reduce energy consumption \cite{Li2023,Zhao2019,Hossain2016,Xia2020Easy,Fan2022}. Owing to the high energy in the range of solar spectrum and transparency of the atmospheric window in the wavelength range of 8 – 14 $\mu$$m$ \cite{elder1953infrared}, there has been a significant effort in developing solar absorbers and radiative coolers using different strategies such as photonic structures \cite{Lee2023,Heo2022,Wang2020,Zhang2021,Ly2022}, metamaterials \cite{Hossain2015,Huang2018,Mishra2021}, and energy-saving paints \cite{Mandal2020,Felicelli2022,Li2021Ultrawhite}. However, most heating/cooling devices designed recently are single-function and non-adjustable, rendering them unsuitable for mid-latitude regions where temperatures fluctuate drastically throughout the year. Therefore, a smart system of passive temperature adjustment is urgently needed.

Here, we propose a temperature-dependence passive solar-absorber and radiative-cooler (TDPSR) system, developed through Bayesian optimization. Using the phase change material n-octadecane \cite{Su2022}, such a system can switch between solar absorbing and radiative cooling modes within a comfortable temperature range, without any additional energy input for switching. Our structure can achieve an average solar absorption of over 0.85 in solar absorbing mode and average emissivity in the atmospheric window of over 0.8 in the radiative mode. The presented results explore new functionalities of the combination and applications of solar absorbing and radiative cooling, enabling automated regulation and optimization between the two modes, and could potentially lead to significant reduction in energy consumption and enhancement of thermal comfort, applicable to a wide range of uses including vehicles, buildings, textiles, and smart thermal regulations.

\section{Results and Discussion}
\label{sec:headings}
\subsection{Design of TDPSR}
The working principle of the TDPSR is presented in Fig. 1B, whereas the expected spectral behaviors of its absorptance and emissivity are shown in Fig. 1A and Fig. 1C. We aim for the TDPSR to facilitate cooling by dissipating heat during hot summers, and to absorb maximum thermal energy from the sun during cold winters to increase indoor temperatures. By using the phase change material n-octadecane with a phase transition temperature at $T_p$ (22.7-31.8°C), slightly above room temperature \cite{zhang2021preparation}, TDPSR can easily maintain a comfortable surrounding temperature by switching sun absorbing mode and radiative cooling mode.

TDPSR with a multilayer structure (Fig. 1D) is designed to meet the spectral requirements for smart dynamic switching between a solar absorber and a radiative cooler. $HfO_2$ and n-octadecane are alternately deposited to manufacture the Epsilon-Near-Zero Metamaterials (ENZM), which are then alternately layered with $SiO_2$ on top of a 200 nm of silver ($Ag$) layer. Although we explored other material combinations and optimized them, the structure presented here proved to be the most effective (\hyperref[SI. Section I]{\textit{SI, section I, Fig. S1-Fig. S3}}). The n-octadecane ﬁlm is essential for the smart integration of solar absorber and radiative cooler due to its different optical properties before and after the phase change. Variables used in the machine learning design are shown in Fig. 1E. By altering the combination of these variables, the optical properties of the TDPSR can be optimized. We employ Bayesian optimization to identify the variable combination with the best optical properties.

\begin{figure}[h]
    \centering
    \begin{subfigure}[b]{0.3\textwidth}
        \centering
        \includegraphics[width=\textwidth]{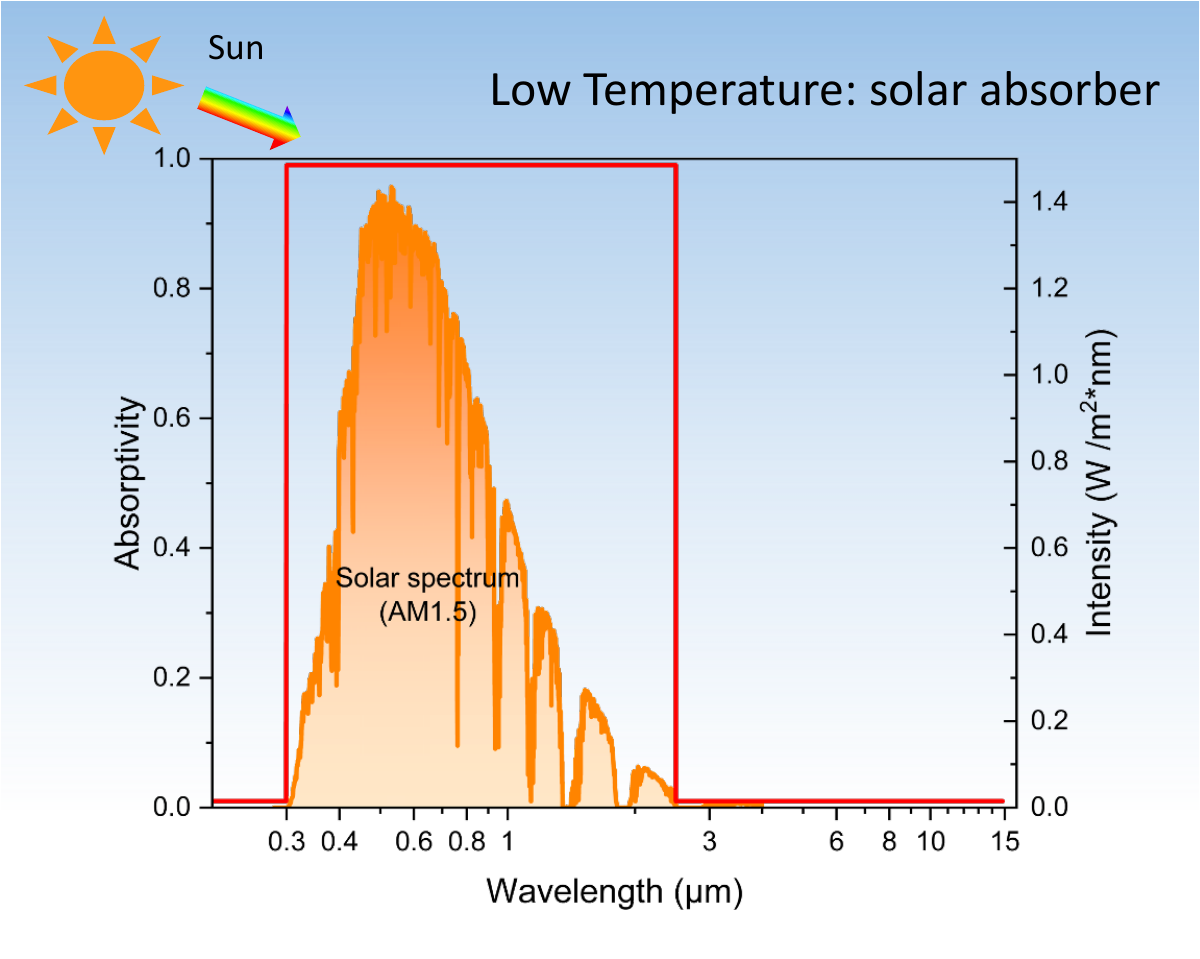}
        \caption*{A}
        \label{fig:a}
    \end{subfigure}
    \hspace{0.01\textwidth}
    \begin{subfigure}[b]{0.3\textwidth}
        \centering
        \includegraphics[width=\textwidth]{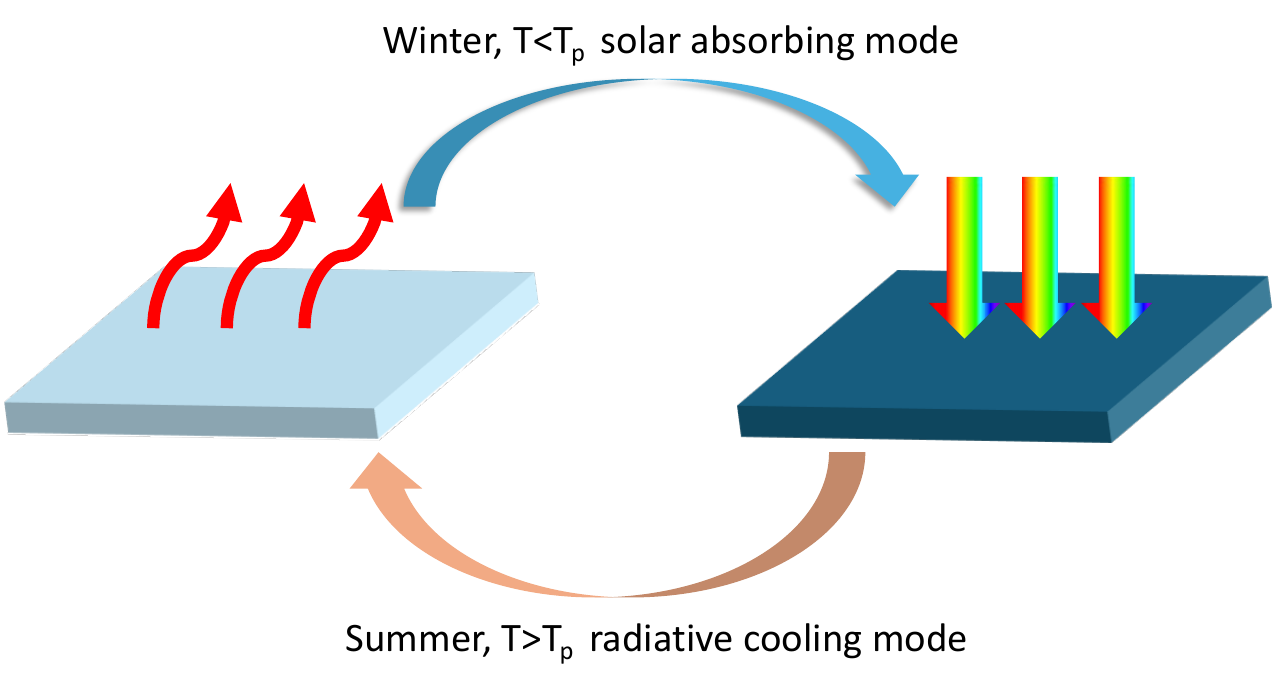}
        \caption*{B}
        \label{fig:b}
    \end{subfigure}
    \hspace{0.01\textwidth}
    \begin{subfigure}[b]{0.3\textwidth}
        \centering
        \includegraphics[width=\textwidth]{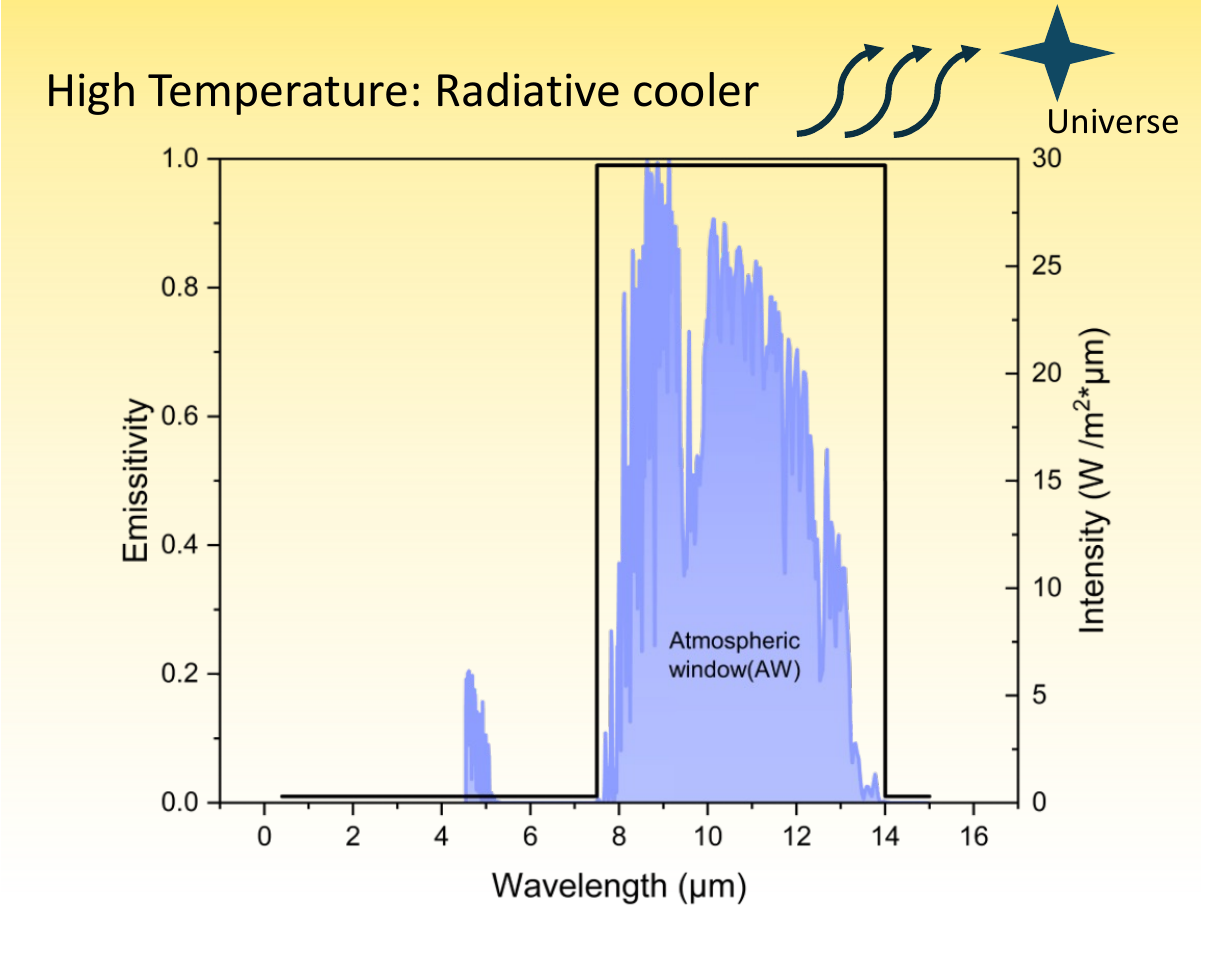}
        \caption*{C}
        \label{fig:c}
    \end{subfigure}

    \begin{subfigure}[b]{0.4\textwidth}
        \centering
        \includegraphics[width=0.8\textwidth]{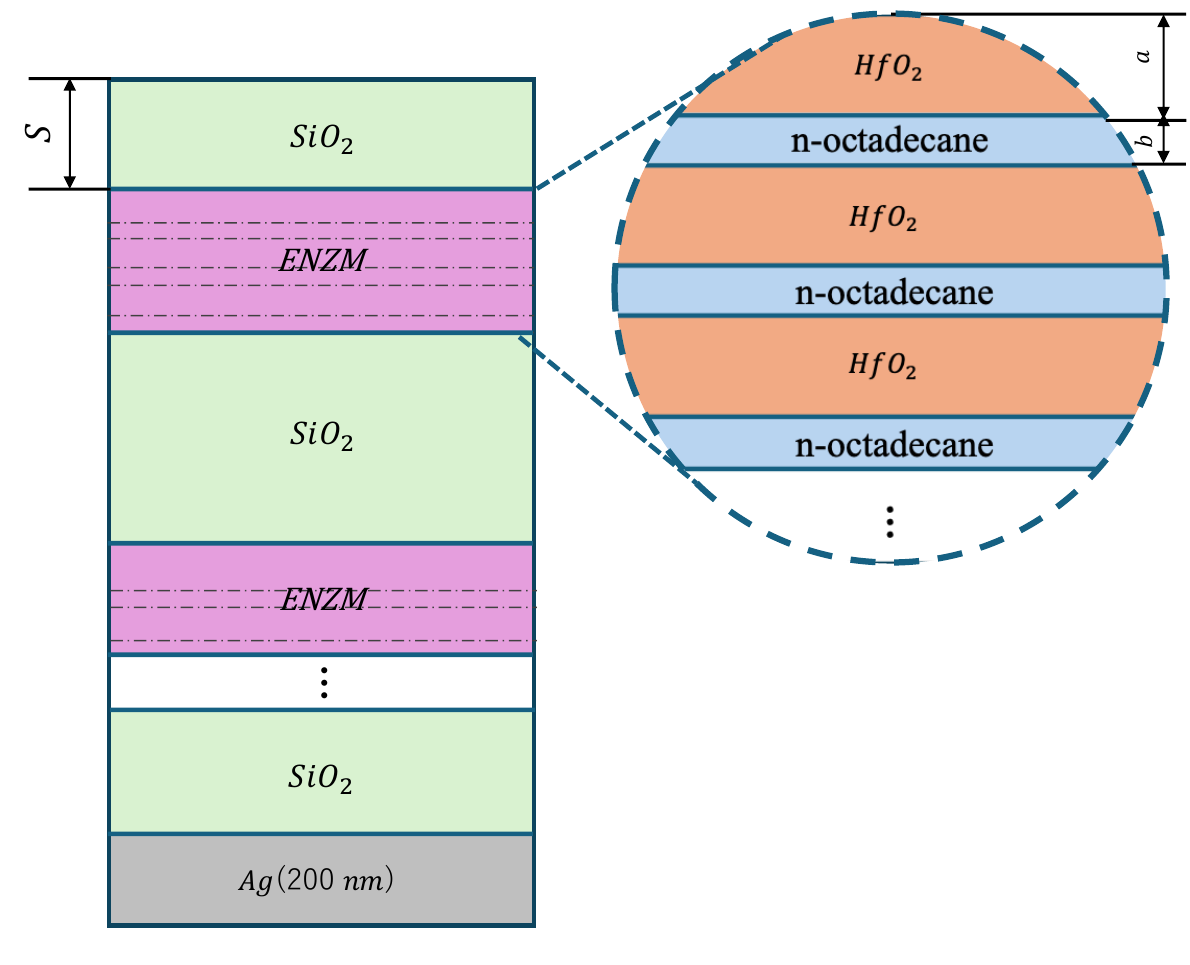}
        \caption*{D}
        \label{fig:d}
    \end{subfigure}
    \hspace{0.05\textwidth}
    \begin{subfigure}[b]{0.4\textwidth}
        \centering
        \includegraphics[width=0.8\textwidth]{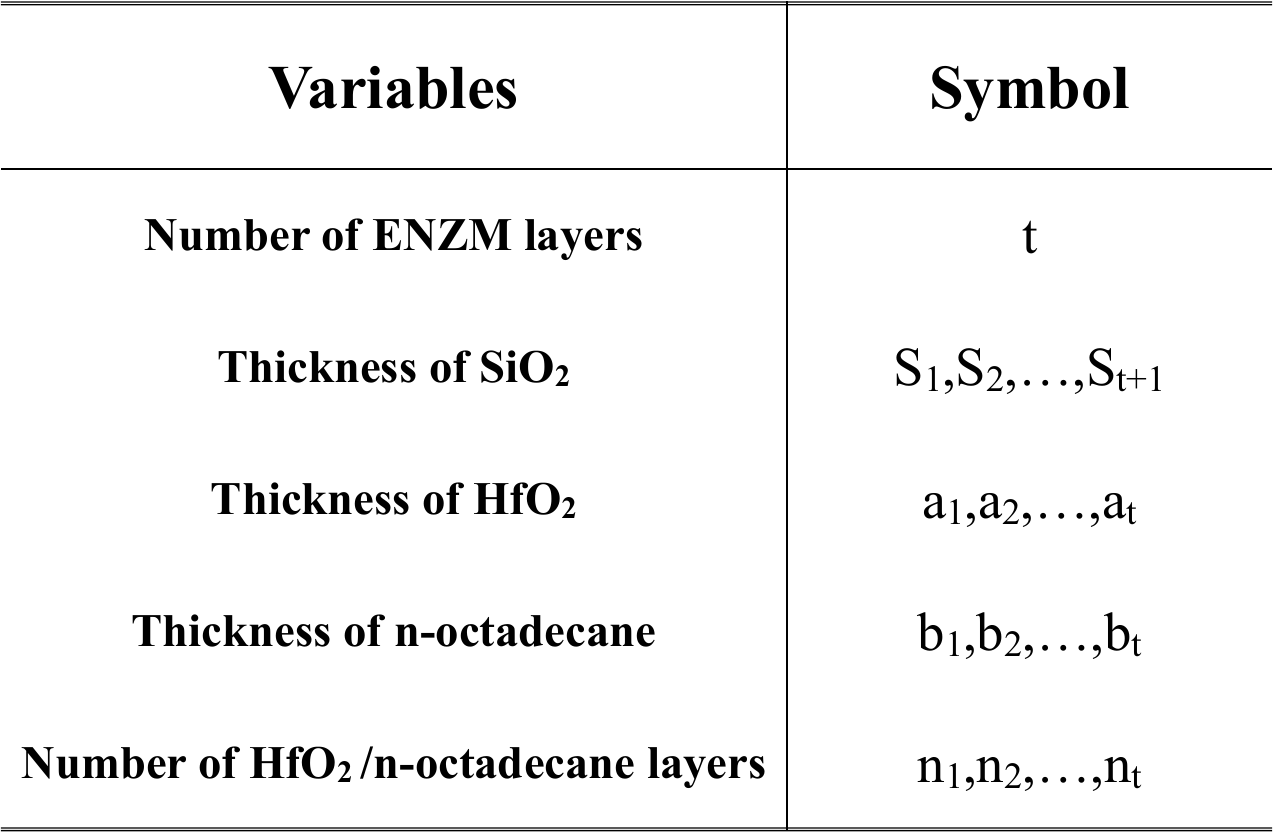}
        \caption*{E}
        \label{fig:e}
    \end{subfigure}
    \caption*{
    Fig. 1: Concept of dynamic integration of solar absorber and radiative cooler based on the TDPSR. (B) Working principle of the TDPSR for the smart integration of solar absorber and radiative cooler. $T$ represents the temperature of the TDPSR, and $T_p$ is the phase change temperature of the ENZM layer. In summer, $T > T_p$, the TDPSR functions as a radiative cooler. During the winter, $T < T_p$, the TDPRS works as a solar absorber. (A) and (C) show the ideal spectrum requirement for the material to dynamically integrate solar absorption and radiative cooling at high and low temperatures. AM1.5: normalized AM1.5 solar spectrum; AW: atmospheric window. (D) Structure of TDPSR. It consists of a periodic $HfO_2$ and n-octadecane metamaterial (ENZM) and a periodic $SiO_2$-ENZM structure, all on top of a 200 $nm$ layer of silver ($Ag$). (E) Variables and symbols used for Bayesian optimization.}
    \label{fig:total}
\end{figure}

The diagram of our Bayesian optimization is shown in Fig. 2A. Our optimization is firstly separated into 5 groups with different ENZM layers number t. The thickness of $SiO_2$, the thickness, and the number of layers of $HfO_2$ and n-octadecane in ENZM are used as parameters for the Bayesian optimization.

According to the expected spectral behavior of absorptance and emissivity of TDPSR, the evaluation of the designed multilayered metamaterial employs the following ﬁgure of merit ($FoM$):

\begin{equation}
FoM_{High}=\frac{\int_{\lambda_3}^{\lambda_4}{\varepsilon I_ed\lambda}}{\int_{\lambda_3}^{\lambda_4}{I_ed\lambda}}-\frac{\int_{\lambda_1}^{\lambda_2}{\varepsilon I_{ra}d\lambda}}{\int_{\lambda_1}^{\lambda_2}{I_{ra}d\lambda}}
\end{equation}

\begin{equation}
FoM_{Low}=\frac{\int_{\lambda_1}^{\lambda_2}{\alpha I_{ra}d\lambda}}{\int_{\lambda_1}^{\lambda_2}{I_{ra}d\lambda}}-\frac{\int_{\lambda_3}^{\lambda_4}{\alpha I_ed\lambda}}{\int_{\lambda_3}^{\lambda_4}{I_ed\lambda}}
\end{equation}

\begin{equation}
FoM_{Op}\ =\frac{1}{2}(FoM_{High}+FoM_{Low})
\end{equation}

where $\varepsilon$ is the emissivity at high temperature, $\alpha$ is the absorption at low temperature, $I_e$ is the intensity of blackbody’s radiation which can be emitted into outer space at 310 K, $\lambda$ is the wavelength, and $I_{ra}$ is the solar radiation intensity reaching the earth’s surface. In our optimization, the range ($\lambda_1$, $\lambda_2$) from 0.4 to 2.5 $\mu$$m$ corresponds to the solar spectrum range (AM1.5). The range ($\lambda_3$, $\lambda_4$) from 8 to 14 $\mu$$m$ corresponds to the atmospheric window (AW). Comprehensive $FoM_{Op}$ will be used to evaluate the performance of TDPSR at both high and low temperatures.

\begin{figure}[h]
    \centering
    \begin{subfigure}[b]{0.3\textwidth}
        \centering
        \includegraphics[width=0.7\textwidth]{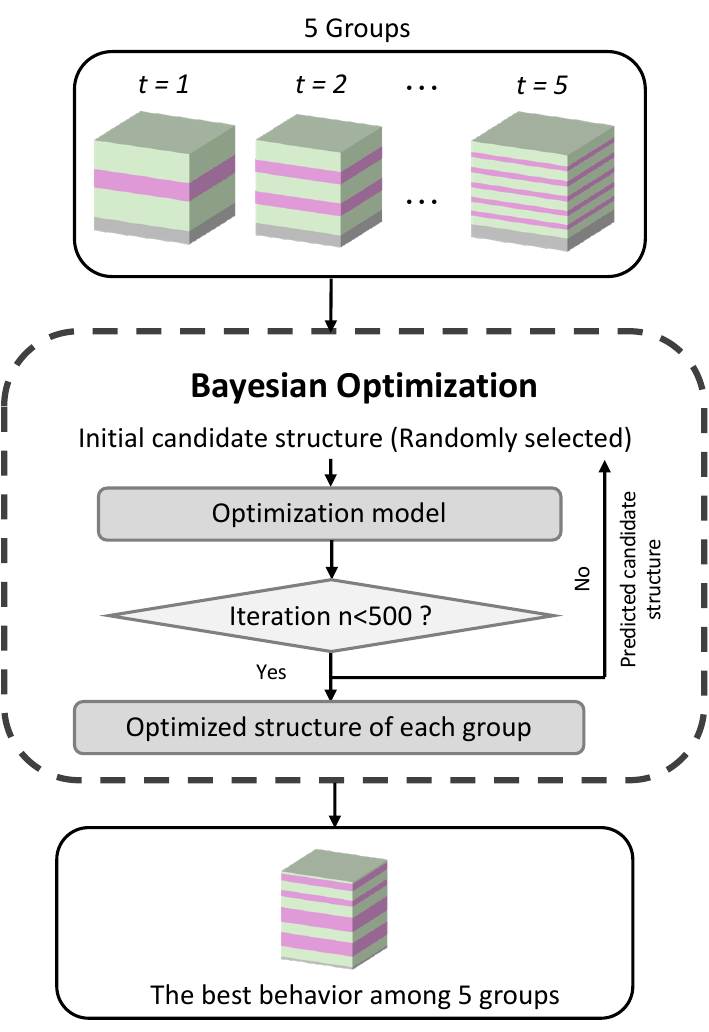}
        \caption*{A}
        \label{fig:a}
    \end{subfigure}
    \hspace{0pt}
    \begin{subfigure}[b]{0.3\textwidth}
        \centering
        \includegraphics[width=1.2\textwidth]{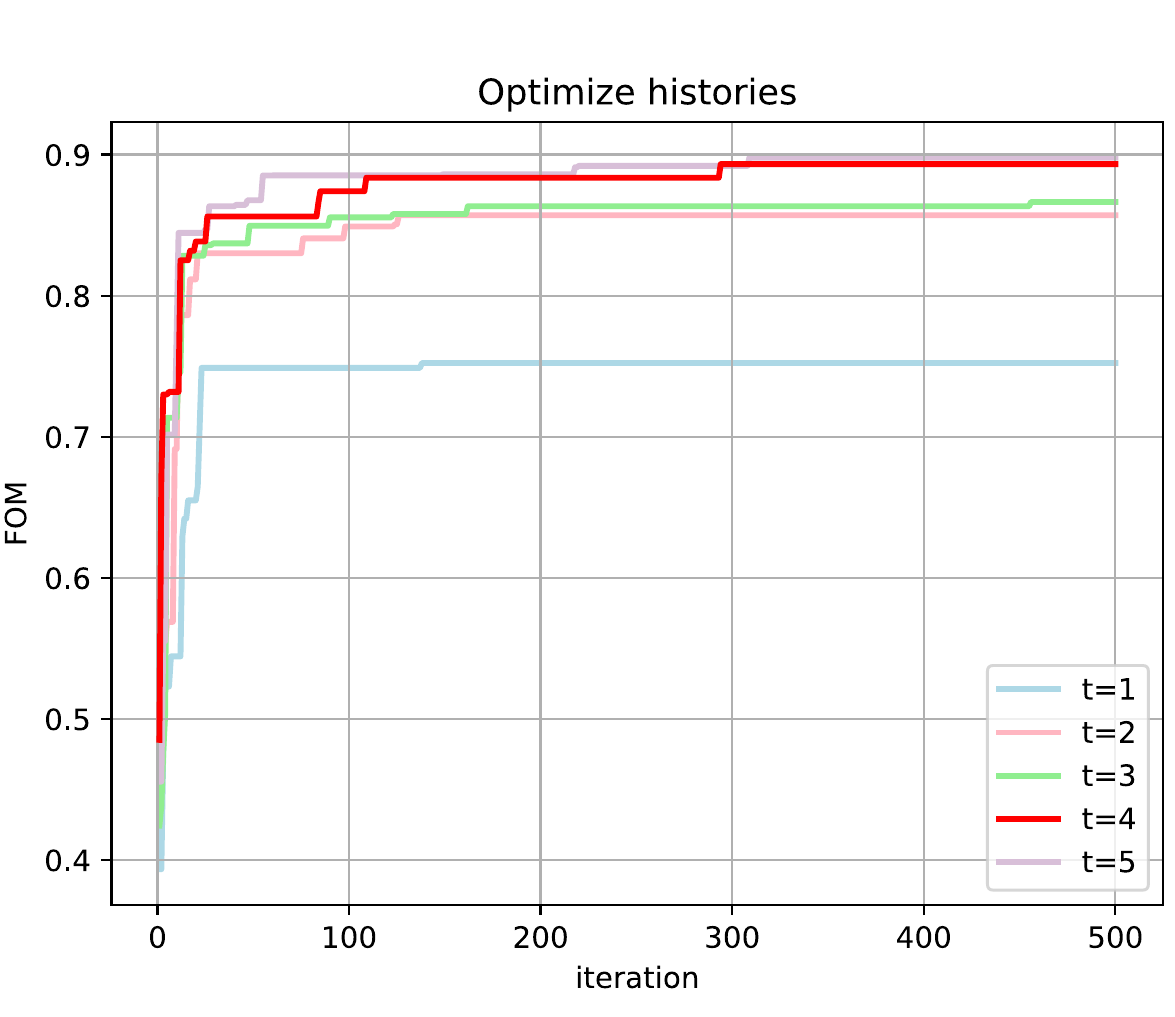}
        \caption*{B}
        \label{fig:b}
    \end{subfigure}
    \hspace{0.06\textwidth}
    \begin{subfigure}[b]{0.3\textwidth}
        \centering
        \raisebox{0.4cm}{ 
            \includegraphics[width=0.6\textwidth]{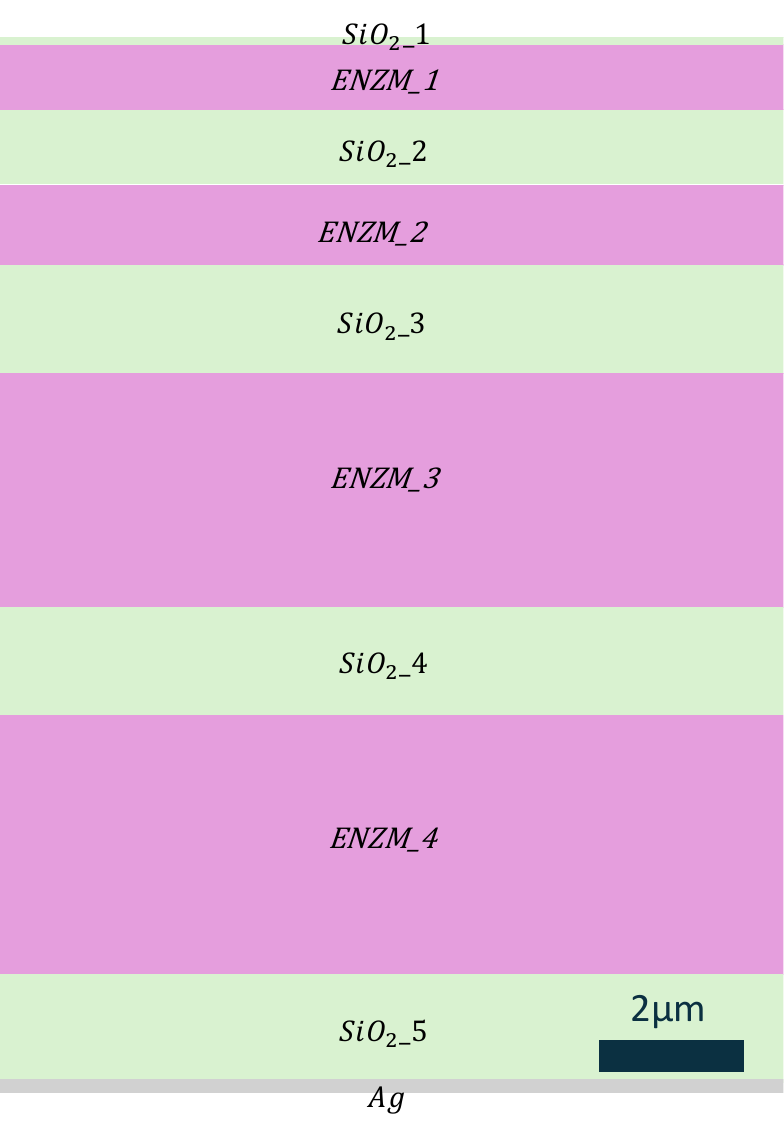}
        }
        \caption*{C}
        \label{fig:c}
    \end{subfigure}
    \caption*{
    Fig. 2: (A) Schematic of the optimization method with Bayesian optimization: The number of ENZM layers in TDPSR (denoted as $t$) serves as the basis for grouping. All the candidates are divided into 5 groups. The parameters for Bayesian optimization include the thickness of $SiO_2$, the thickness and number of layers of $HfO_2$ and n-octadecane in ENZM. After 500 iterations, the parameter combination with the best optical performance of each group was identified. The $FoM$ of each group was then compared, considering the total thickness and the optical performance in invalid wavelength ranges (i.e., TDPSR should not have excessive absorptivity/emissivity in the 3-8 $\mu$$m$ region, which is neither AM1.5 nor AW) to find the optimal TDPSR structure. (B) Histories of the $FoM$s of 5 groups. ENZM with 4 and 5 layers have similar highest $FoM$. (C) Structural diagram of TDPSR with four-layer ENZM. As it is easier to manufacture, it is chosen as the optimal structure.}
    \label{fig:total}
\end{figure}

\begin{figure}[h]
    \centering
    \caption*{\textbf{Table 1: Parameters of the optimized structure obtained by Bayesian optimization.}}
    \label{tab:parameters}
    \includegraphics[width=0.7\textwidth]{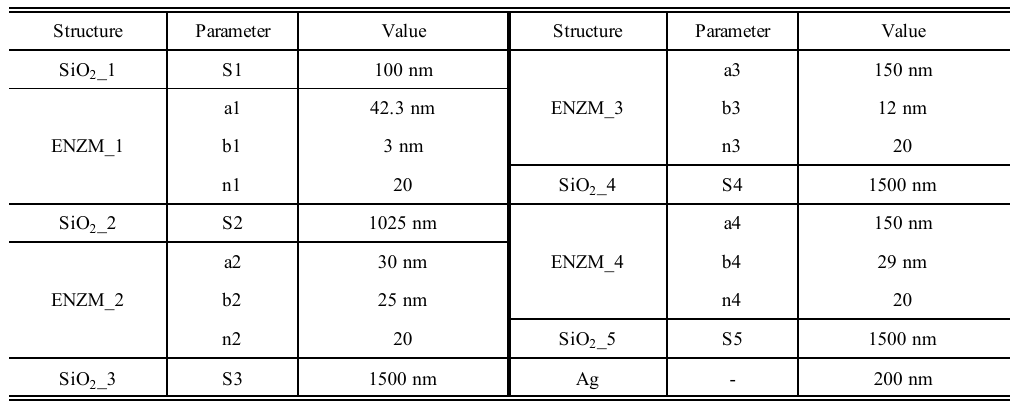} 
\end{figure}

As shown in Fig. 2A, suppose that $FoM$s of n candidates are initially calculated, and we are going to select the next ones to calculate. A Bayesian optimization model will learn from n pairs of input parameters and $FoM$s (i.e. training examples) to get the distribution of $FoM$s. The optimal $FoM$ and its corresponding parameter combination are identified within this distribution and then added to the training examples. By repetition of this procedure, the distribution of $FoM$ obtained through Bayesian optimization will increasingly approximate the true distribution, allowing the optimization to gradually approach the maximum value of $FoM$. By using Bayesian optimization, the optimized structure can be found quickly.

The variations of the maximum $FoM$ with respect to the number of iterations are shown in Fig. 2B. We examine the cases of 5 groups and find that TDPSR with $t$ = 4 and $t$ = 5 have similar high $FoM$s. The structure corresponding to $t$ = 4 is selected due to its ease of manufacturing (\hyperref[SI. Section II]{\textit{SI, section II, Table S1}}). The maximum $FoM$ of each group could be reached within 300 iterations, demonstrating the efficiency of the implemented Bayesian optimization.

\begin{figure}[h]
    \centering
    \begin{subfigure}[b]{0.35\textwidth}
        \centering
        \includegraphics[width=\textwidth]{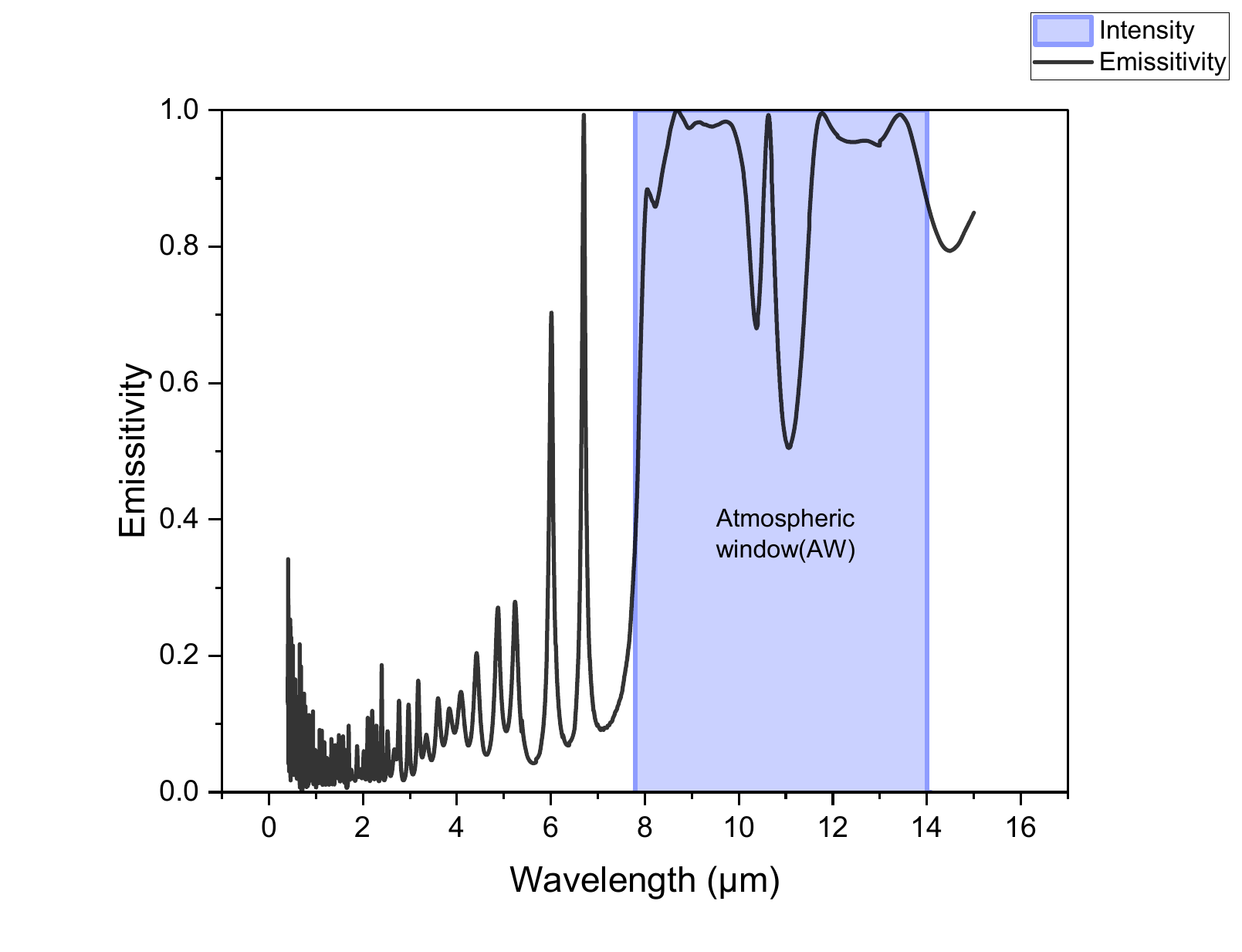}
        \caption*{A}
        \label{fig:a}
    \end{subfigure}
    \hspace{0.01\textwidth}
    \begin{subfigure}[b]{0.35\textwidth}
        \centering
        \includegraphics[width=\textwidth]{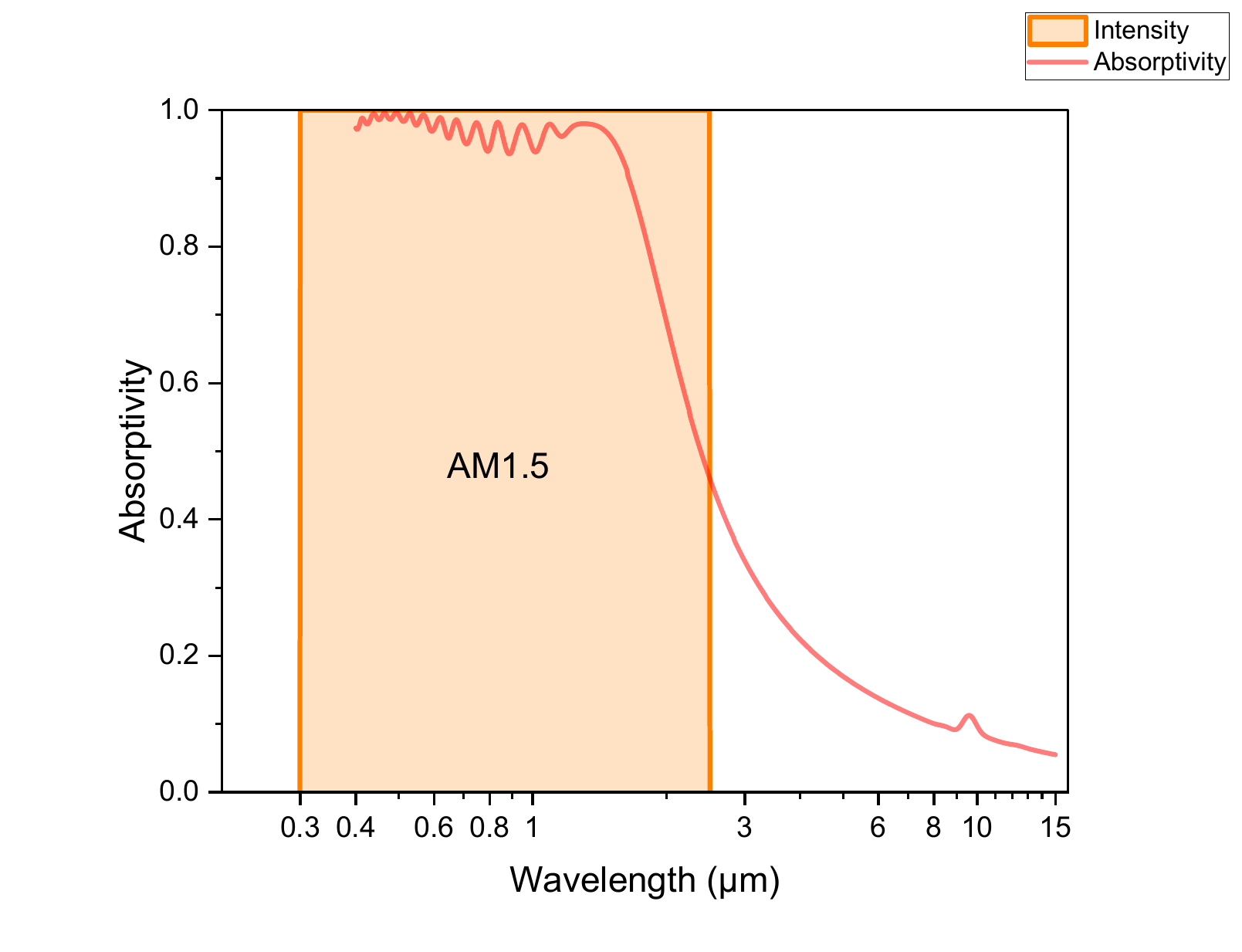}
        \caption*{B}
        \label{fig:b}
    \end{subfigure}
    
    \begin{subfigure}[b]{0.35\textwidth}
        \centering
        \includegraphics[width=\textwidth]{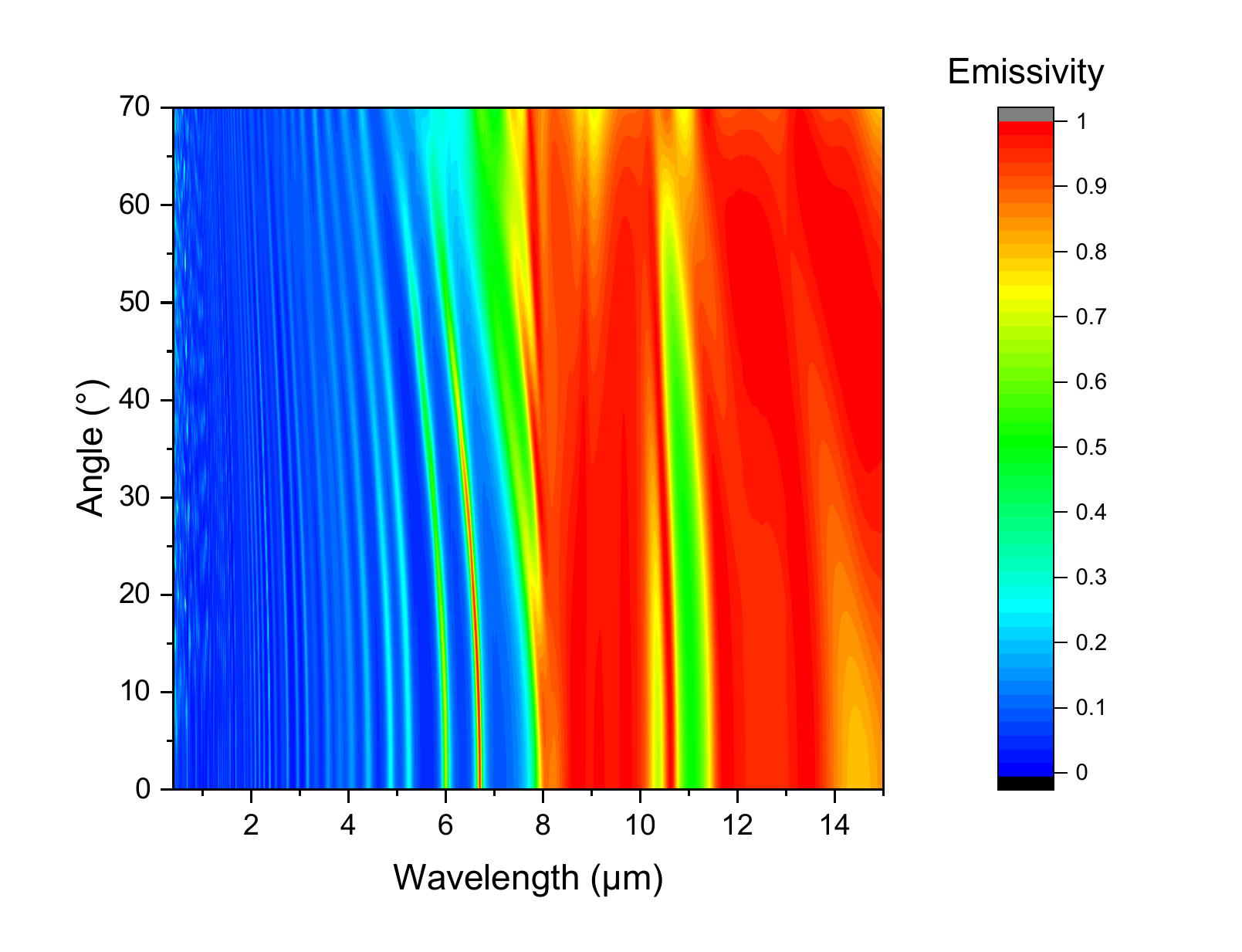}
        \caption*{C}
        \label{fig:a}
    \end{subfigure}
    \hspace{0.01\textwidth}
    \begin{subfigure}[b]{0.35\textwidth}
        \centering
        \includegraphics[width=\textwidth]{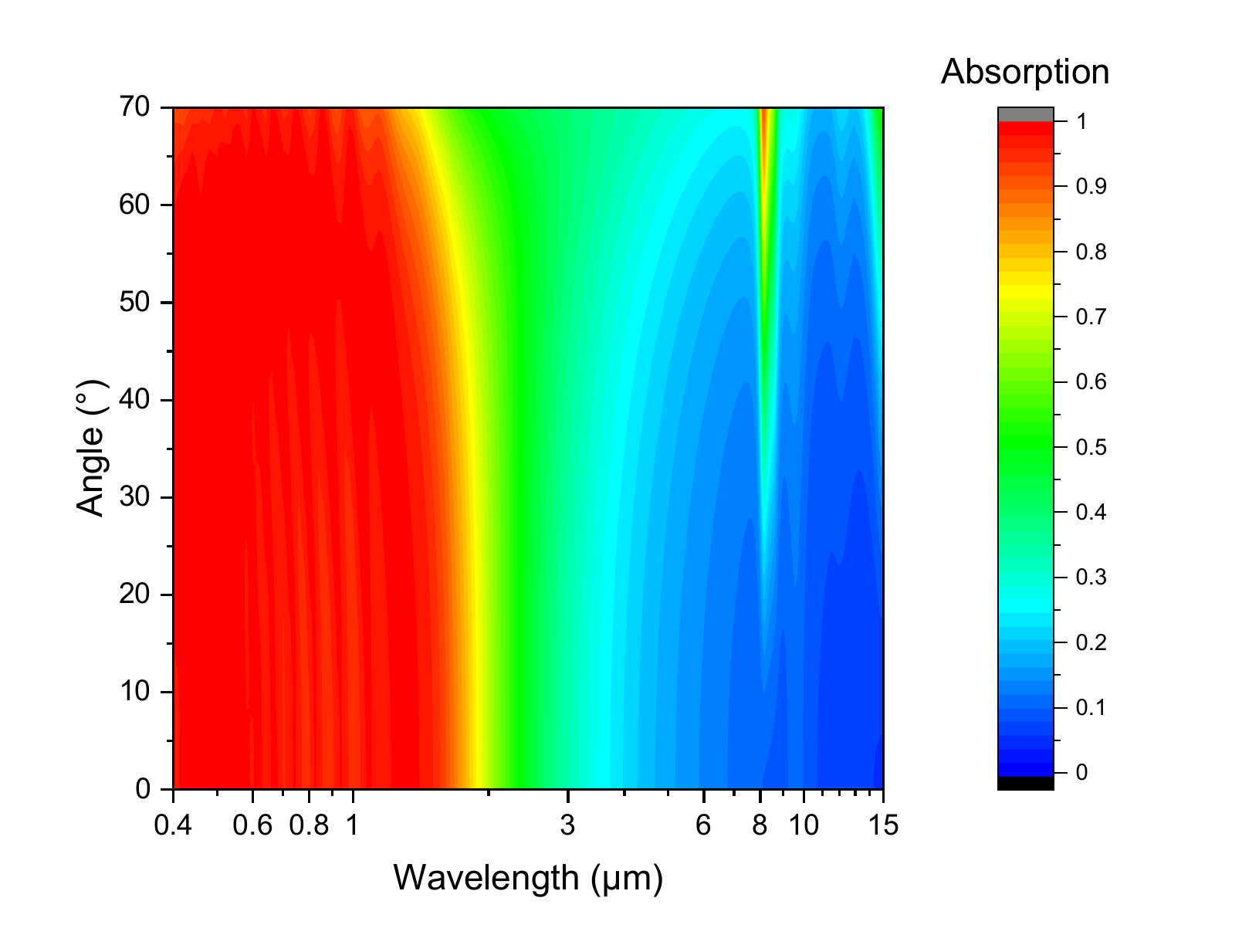}
        \caption*{D}
        \label{fig:b}
    \end{subfigure}
    \caption*{Fig. 3: Spectral absorption/emissivity of TDPSR. (A) Emissivity of TDPSR from the UV to the MIR band (above the phase change temperature $T_p$ of n-octadecane). (B) Absorption of TDPSR from the UV to the MIR band at low temperatures (below phase change temperature $T_p$ of n-octadecane). (C) Angle-wavelength-emissivity color map of TDPSR at high temperatures, showing high emissivity in the atmospheric window over a wide range of angles. (D) Angle-wavelength-emissivity color map of TDPSR at low temperatures shows high absorption for AM1.5 across a wide range of angles.}
    \label{fig:total}
\end{figure}

\subsection{Optical properties of TDPSR}
 Periodic structured ENZM functions as the smart switch for temperature regulation, enabling temperature-dependent optical properties of TDPSR (\hyperref[SI. Section III]{\textit{SI, section III, Fig. S4}}). The model of AW \cite{AM1.5} and AM1.5 \cite{AW} are used to demonstrate the TDPSR’s energy emitting and absorbing behavior. When the temperature of the TDPSR is higher than $T_p$, the device exhibited substantially high emissions within the atmospheric window as shown in Fig. 3A, with a $FoM$ over 0.879. When $T < T_p$, the n-octadecane layer converted to the metallic rutile phase, resulting in a significant drop in emissivity within AW and an increase in absorption within AM1.5, as shown in Fig. 3B, with $FoM$ over 0.967.

The angular-dependent emissivity/absorption was also investigated. The solar zenith angle at mid-latitudes (30°- 45°) at noon (11 am - 2 pm) throughout the year was simulated (\hyperref[SI. Section IV]{\textit{SI, section IV, Fig. S5}}) and found to fluctuate between 0 and 70° \cite{Abdulameer2015,Li2021Optimal}. Fig. 3C and Fig. 3D analyze the angular dependence of the absorption of the metamaterial. As shown in these figures, the calculated absorption/emissivity at a given wavelength in both AW and AM1.5 spectrum is largely independent of the angle of incidence up to 70°. Therefore, the optimized TDPSR performs similarly across a wide range of angles, thus offering clear advantages over an omnidirectional emitter because it allows extracting and transmitting a much larger radiative power.

\subsection{Thermal effects of TDPSR}
The evaluation is conducted on a sunny day in both summer and winter. To begin the analysis, TDPSR at temperature $T$, whose radiative properties are described by a spectral and angular emissivity or angular absorption $\varepsilon (\lambda, \theta)$ is considered. Then a heat transfer model is used to predict the stagnation temperature and net output energy of TDPSR. The structure is expected to be exposed to a clear sky subject to solar irradiance, while also being affected by the atmospheric irradiance corresponding to ambient temperature $T_{amb}$.

\begin{figure}[h]
    \centering
    \begin{subfigure}[b]{0.4\textwidth}
        \centering
        \includegraphics[width=\textwidth]{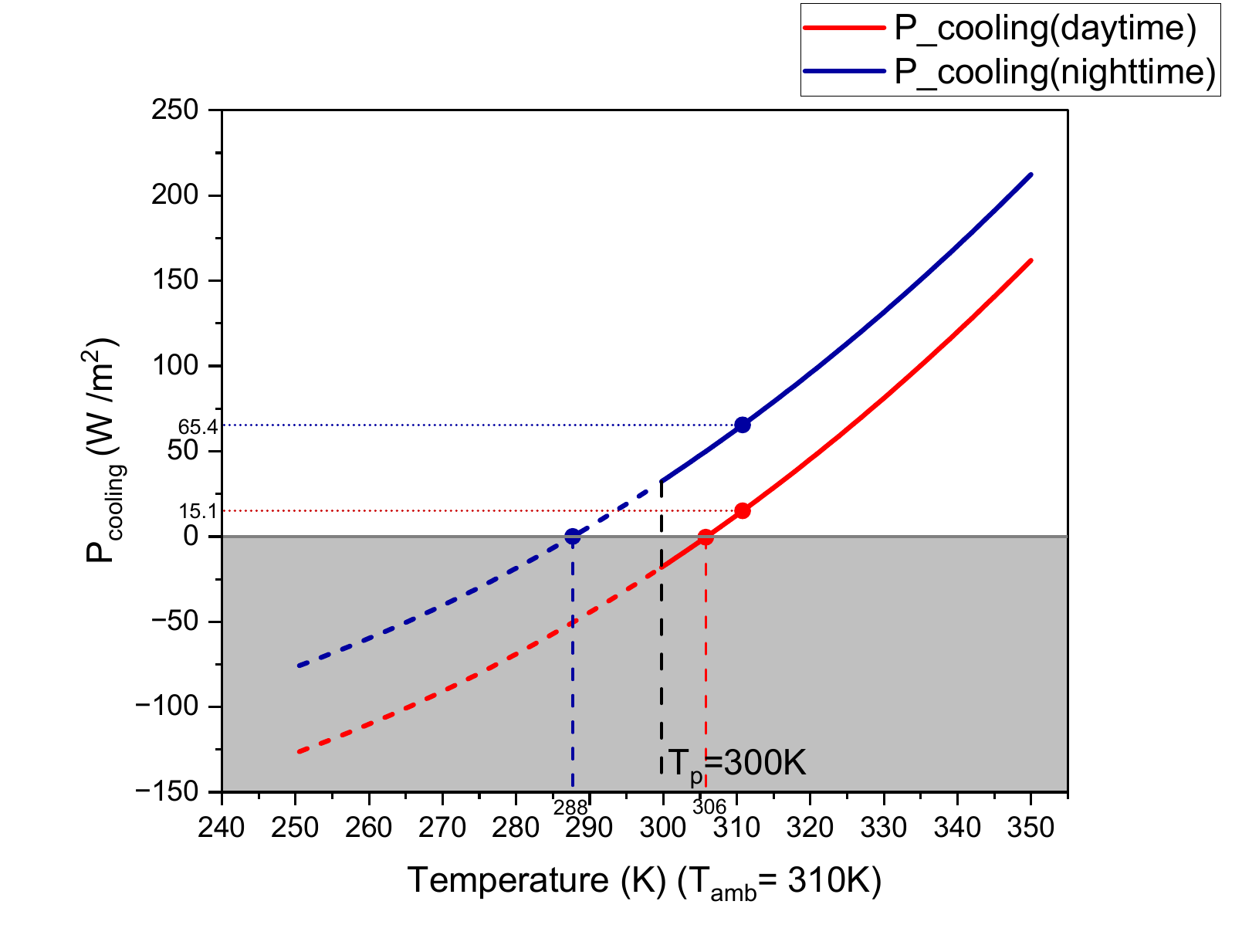}
        \caption*{A}
        \label{fig:a}
    \end{subfigure}
    \hspace{0.01\textwidth}
    \begin{subfigure}[b]{0.4\textwidth}
        \centering
        \includegraphics[width=\textwidth]{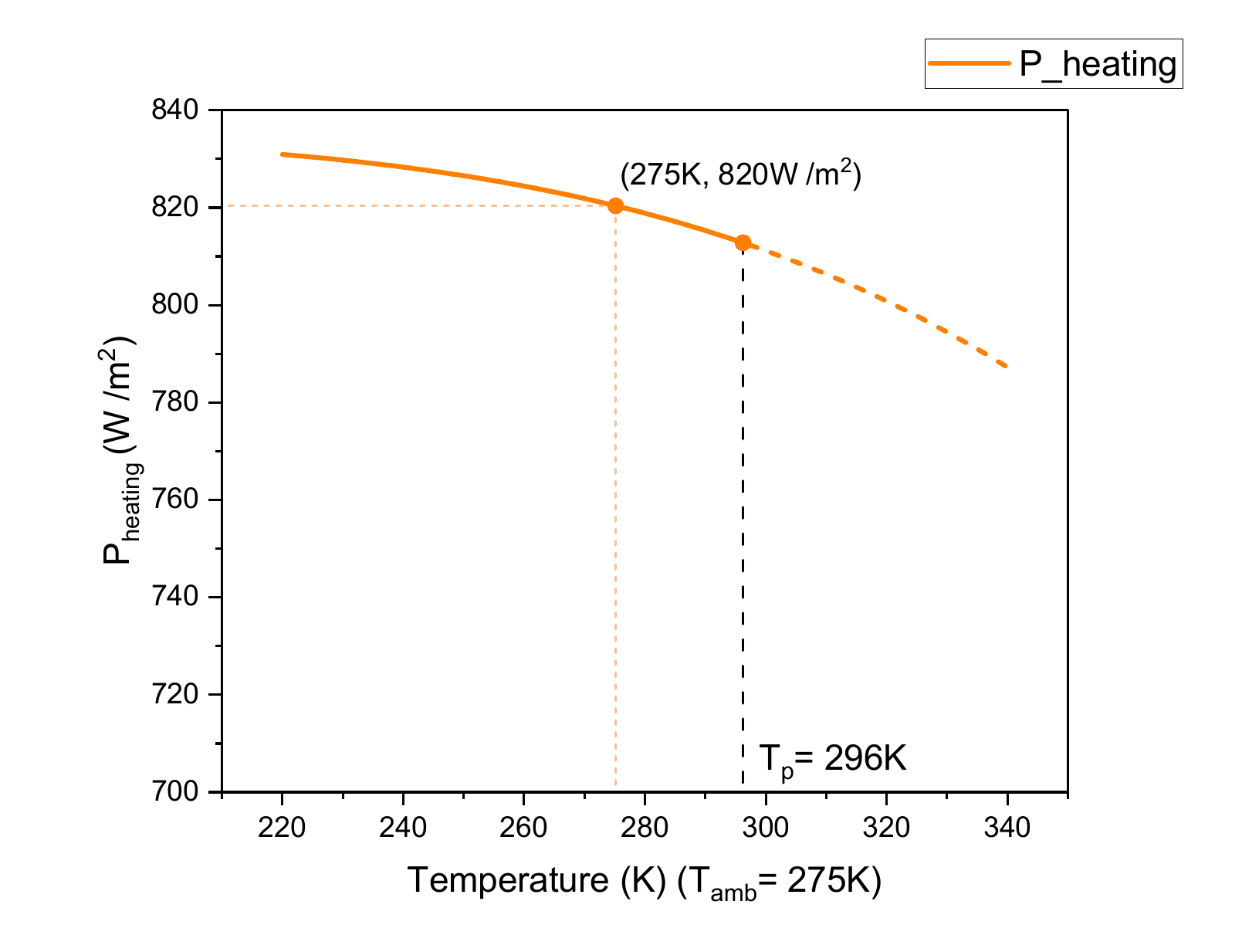}
        \caption*{B}
        \label{fig:b}
    \end{subfigure}
    \caption*{
    Fig. 4 Thermodynamic performance of TDPSR at high and low temperatures. (A) The red line represents the cooling power at nighttime for an ambient temperature of 310 $K$ (36.85 °C), reaching an equilibrium temperature of 288 $K$ (14.85 °C) with a cooling power of 65.4 $W/m^2$ at $T_amb$. The blue line represents the power of cooling in the daytime for an ambient temperature of 310 $K$, reaching an equilibrium temperature of 306 $K$ (32.85 °C) with a cooling power of 15.1 $W/m^2$ at Tamb. (B) The heating power at the ambient temperature of 275 $K$ (1.85 °C). TDPSR can reach a very high heating power Pheating of 820 $W/m^2$ from the sun radiation.}
    \label{fig:total}
\end{figure}

Using the heat transfer model (\hyperref[SI. Section V]{\textit{SI, section V, Fig. S6}}), we plot $P_{cooling}$ ($T$) given $T_{amb} = 310 K$ and $P_{heating}$ ($T$) given $T_{amb} = 275K$ for TDPSR, calculated via hemispheric integration of the structure’s emissivity. As shown in Fig. 4A, $P_{cooling}$ during the daytime is shown along with the nighttime cooling power $P_{cooling}$($nighttime$), achieved by setting $P_{sun}$ = 0. Our designed structure exhibits a remarkably low nighttime equilibrium temperature of 288 $K$, and a daytime equilibrium temperature of 306 $K$. It furthermore has a cooling power of 65.4 $W/m^2$ at $T_{amb}$ at nighttime and 15.1 $W/m^2$ at $T_{amb}$ in the daytime. As shown in Fig. 4B, at low temperatures with $T_{amb} = 275 K$, it can achieve an especially high heating power of 820 $W/m^2$ from the sun. Thus, TDPSR strikes a good balance between heating and cooling, demonstrating excellent thermal properties at both high and low temperatures.

\section{Summary}
A temperature-dependence passive solar-absorber and radiative-cooler (TDPSR) system, incorporating $SiO_2$ and Epsilon-Near-Zero Metamaterials (ENZM) with a periodic $HfO_2$ and n-octadecane structure, is designed using machine learning. The TDPSR exhibits strong sunlight absorption with a figure of merit ($FoM$) over 0.967 at low temperature and selective high emissivity with a $FoM$ over 0.879 within the atmospheric window. Such an excellent optical properties can be maintained not only under direct sunlight but also across a wide angular range from 0° to 70°, demonstrating remarkable practicality in the mid-latitude region. Simulation of thermal performance of TDPSR at high and low temperatures also demonstrates effective cooling/heating efficiency. In summary, this work presents a method to dynamically explore the thermodynamic potential of the hot sun and the cold universe through the smart temperature-dependent integration of a solar absorber and radiative cooler. This approach not only enables new technological capabilities but also enhances our understanding of designing metamaterial structures through the use of machine learning.

\section*{Supplementary Information}
\subsection*{I. Investigation into the effect of the layer combinations }
\label{SI. Section I}
Calculation results for different layer combinations are shown below: (Step 1) exchanging Layer-$HfO_2$ to Layer-$SiO_2$ in all layers and (Step 2) changing Layer-$HfO_2$ to Layer-$SiO_2$.

\begin{figure}[h]
    \centering
    \begin{subfigure}[b]{0.7\textwidth}
        \centering
        \includegraphics[width=\textwidth]{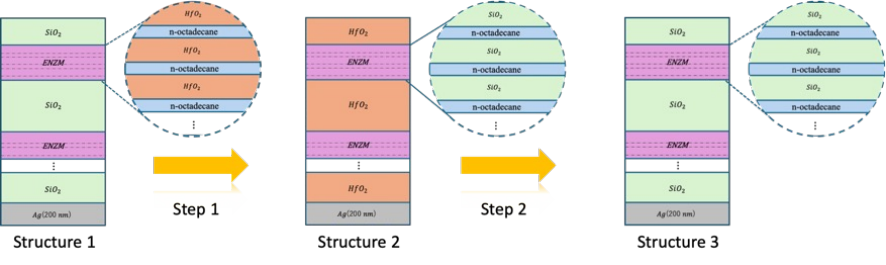}
        \caption*{A}
        \label{fig:a}
    \end{subfigure}

    \begin{subfigure}[b]{0.35\textwidth}
        \centering
        \includegraphics[width=\textwidth]{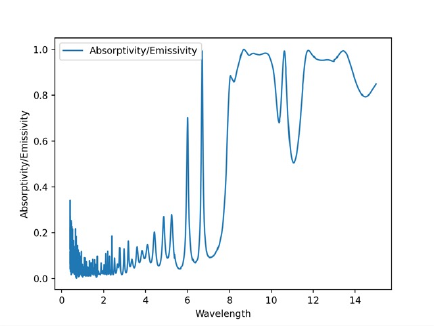}
        \caption*{B}
        \label{fig:a}
    \end{subfigure}
    \hspace{0.01\textwidth}
    \begin{subfigure}[b]{0.35\textwidth}
        \centering
        \includegraphics[width=\textwidth]{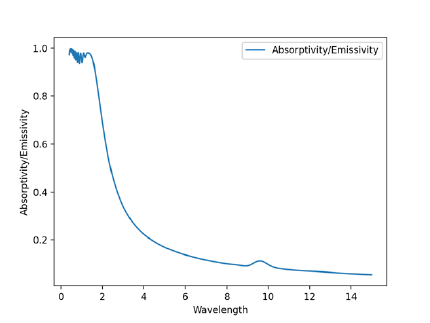}
        \caption*{C}
        \label{fig:b}
    \end{subfigure}
    \caption*{
    Fig. S1: (A) Schematic of TDPSR before changing and after changing. (B) Absorption under high temperature of TDPSR before changing. (C) Absorption under low temperature of TDPSR before changing.}
    \label{fig:total}
\end{figure}

First, the optimal behavior before changing (Structure 1) has been measured with $FoM_{hightemp}$ = 0.858 and $FoM_{lowtemp}$ = 0.967.

\begin{figure}[h]
    \centering
    \begin{subfigure}[b]{0.35\textwidth}
        \centering
        \includegraphics[width=\textwidth]{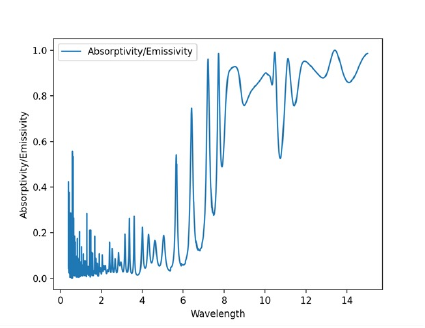}
        \caption*{A}
        \label{fig:a}
    \end{subfigure}
    \hspace{0.01\textwidth}
    \begin{subfigure}[b]{0.35\textwidth}
        \centering
        \includegraphics[width=\textwidth]{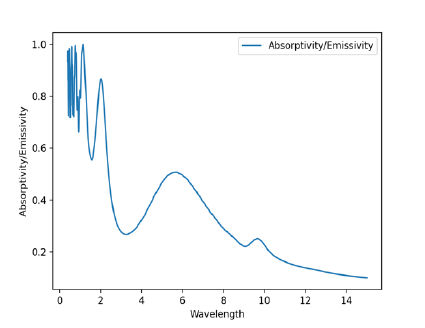}
        \caption*{B}
        \label{fig:b}
    \end{subfigure}
    \caption*{
    Fig. S2: (A) Absorption under high temperature of TDPSR after exchanging Layer-$HfO_2$ and Layer-$SiO_2$. (B) Absorption under low temperature of TDPSR after exchanging Layer-$HfO_2$ and Layer-$SiO_2$.}
    \label{fig:total}
\end{figure}

Then the optimal behavior after exchanging Layer-$HfO_2$ and Layer-$SiO_2$ (Structure 2) has been evaluated with $FoM_{hightemp}$ = 0.848 and $FoM_{lowtemp}$ = 0.832.

\begin{figure}[h]
    \centering
    \begin{subfigure}[b]{0.35\textwidth}
        \centering
        \includegraphics[width=\textwidth]{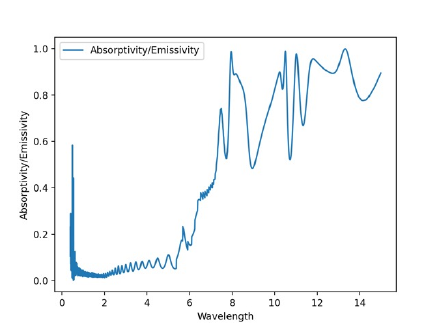}
        \caption*{A}
        \label{fig:a}
    \end{subfigure}
    \hspace{0.01\textwidth}
    \begin{subfigure}[b]{0.35\textwidth}
        \centering
        \includegraphics[width=\textwidth]{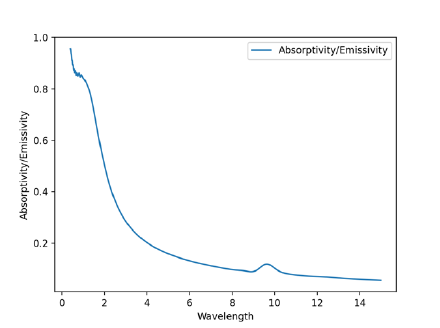}
        \caption*{B}
        \label{fig:b}
    \end{subfigure}
    \caption*{
    Fig. S3: (A) Absorption under high temperature of TDPSR after changing Layer-$HfO_2$ to Layer-$SiO_2$. (B) Absorption under low temperature of TDPSR after changing Layer-$HfO_2$ to Layer-$SiO_2$.}
    \label{fig:total}
\end{figure}

The optimal behavior after changing Layer-$HfO_2$ to Layer-$SiO_2$ (Structure 3) was found to give $FoM_{hightemp}$ = 0.781 and $FoM_{lowtemp}$ = 0.848.

In conclusion, structure 1 is found to give the best results.

\subsection*{II. Average absorption/emissivity and total thickness of TDPSR. }
\label{SI. Section II}
Five groups of TDPRS were optimized separately, and the optimization results are shown in Table S1. In the high-temperature regime, AW (8-14 $\mu$$m$) is the valid range for the calculation, whereas in the low-temperature regime, AM1.5 (0.3-2.5 $\mu$$m$) is the valid range.

The temperature-dependent optical properties of the 4-layer-ENZM and 5-layer-ENZM TDPRS are calculated to select the best behavior for the smart integration of the PT and RC. After comparing the average emissivity/absorption of the optimized structure of ENZM=4 and ENZM=5 in Table S1, the ENZM=4 structure was selected as the best because the extra emissivity/ absorptivity out of the valid range is smaller and the structure is thinner thus easier to manufacture.

\begin{figure}[h]
    \centering
    \caption*{\textbf{Table S1: Average absorption/emissivity of TDPSR}}
    \label{tab:parameters}
    \includegraphics[width=1\textwidth]{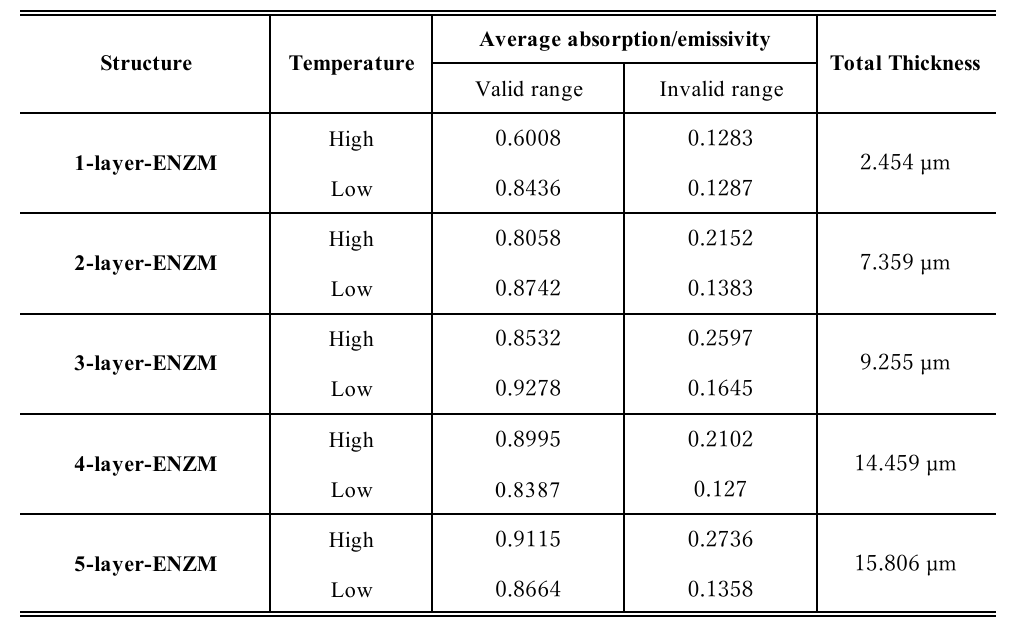} 
\end{figure}

\subsection*{III. Periodic $HfO_2$/ n-octadecane structured ENZM. }
\label{SI. Section III}
In our design, we will utilize a metamaterial structure known as a Hyperbolic Metamaterial (HMM) due to its capability to suppress emissivity in the mid-infrared range \cite{krishnamoorthy2012topological}. This phenomenon needs the material to exhibit metallic properties in one direction and dielectric (insulating) properties in the orthogonal direction. Such anisotropic behavior is not typically found in natural materials at optical frequencies but can be engineered using artificial nanostructured metamaterials. The expression of the iso-frequency surface is given by:

\begin{equation}
\frac{k_x^2+k_y^2}{\varepsilon_\bot}+\frac{k_z^2}{\varepsilon_\parallel}=\ \frac{\omega^2}{c^2}
\end{equation}

We observe that if both $\varepsilon_\bot$ and $\varepsilon_\parallel$ are greater than zero, the iso-frequency surface of the material is an ellipsoid. Conversely, if $\varepsilon_\bot$ is greater than zero and $\varepsilon_\parallel$ is less than zero, the iso-frequency surface becomes a hyperboloid. When all permittivity components are negative, the material behaves as a metal, while if all components are positive, it acts as a dielectric medium. As the wavelength increases, both $\varepsilon_\bot$ and $\varepsilon_\parallel$ decrease, causing the transition from an ellipsoid to a hyperboloid when $\varepsilon_\parallel$ becomes negative. The zero-crossing point can be tailored through material design \cite{shekhar2014hyperbolic}. The mechanism of mid-infrared absorption/emissivity suppression can be illustrated by considering the wave-vector \textbf{k}. In the visible light-near infrared region, the iso-frequency surface of the metamaterial is a closed ellipsoid, thus the wave-vector is confined to the surface of a sphere, allowing normal emission behavior akin to a dielectric. As the incident light wavelength increases, the iso-frequency surface transitions to an open-surface hyperboloid. Consequently, the wave-vector is not confined, leading to an infinite volume. In a vacuum, such large wave-vector waves are evanescent and decay exponentially, thereby suppressing absorption/emissivity \cite{dyachenko2016controlling}.

Drawing inspiration from the design of this metamaterial, we developed a periodic material termed Epsilon-near-zero Metamaterial (ENZM), utilizing the dielectric material $HfO_2$ and the phase-change material n-octadecane. The optical properties of ENZM are modulated by the phase state of n-octadecane. When n-octadecane is in its metallic state, the material exhibits high absorption in the visible to near-infrared regions and suppressed absorption in the mid-infrared region. Conversely, when n-octadecane is in its dielectric state, ENZM behaves as a dielectric. ENZM and $SiO_2$ work together to meet the demands of selective thermal emission in the atmospheric window (AW).

\begin{figure}[h]
    \centering
    \label{tab:parameters}
    \includegraphics[width=1\textwidth]{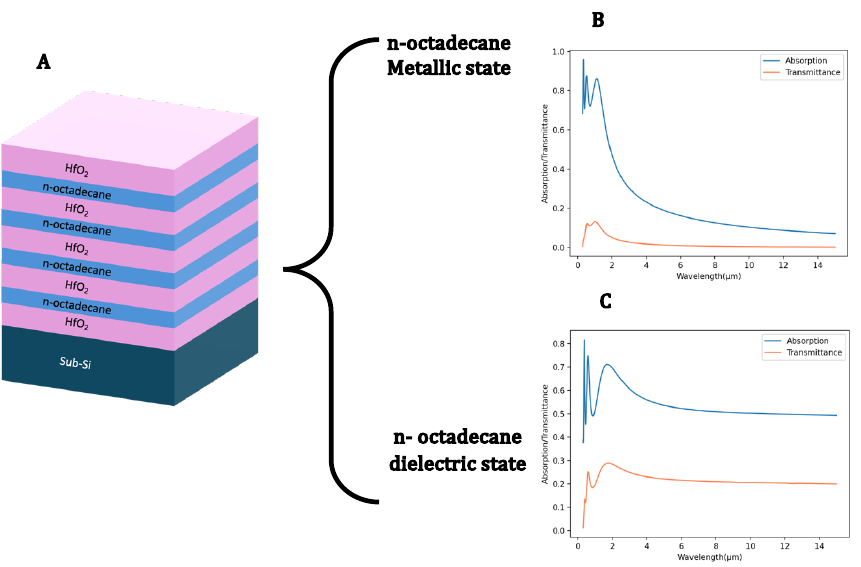} 
    \caption*{Fig. S4: (A) Schematic of $HfO_2$/ n-octadecane periodic metamaterial. (B) Absorption when n-octadecane is in metallic state. (C) Absorption when n-octadecane is in dielectric state.}
\end{figure}

\newpage
\subsection*{IV. Solar zenith angles}
\label{SI. Section IV}
\begin{figure}[h]
    \centering
    \begin{subfigure}[b]{0.35\textwidth}
        \centering
        \includegraphics[width=\textwidth]{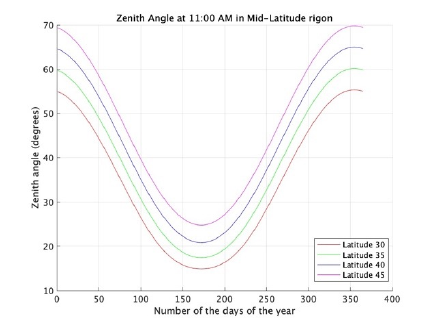}
        \caption*{A}
        \label{fig:a}
    \end{subfigure}
    \hspace{0.01\textwidth}
    \begin{subfigure}[b]{0.35\textwidth}
        \centering
        \includegraphics[width=\textwidth]{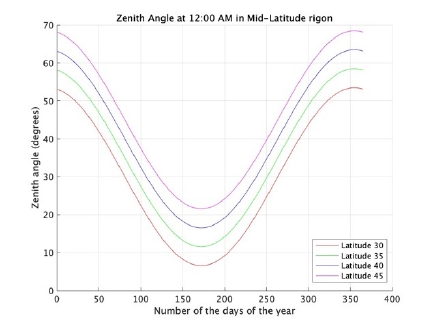}
        \caption*{B}
        \label{fig:b}
    \end{subfigure}

    \begin{subfigure}[b]{0.35\textwidth}
        \centering
        \includegraphics[width=\textwidth]{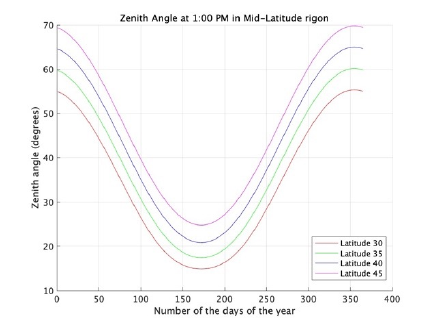}
        \caption*{C}
        \label{fig:a}
    \end{subfigure}
    \hspace{0.01\textwidth}
    \begin{subfigure}[b]{0.35\textwidth}
        \centering
        \includegraphics[width=\textwidth]{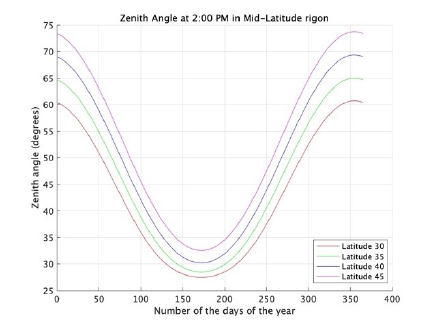}
        \caption*{D}
        \label{fig:b}
    \end{subfigure}

    \caption*{Fig. S5: The solar zenith angle at mid-latitudes (30°- 45°) from 11 am to 2 pm throughout the year\cite{abood2015comprehensive}}
    \label{fig:total}
\end{figure}

\newpage
\subsection*{V. Heat Transfer Model}
\label{SI. Section V}
Using the heat transfer model to measure the thermal performance of TDPSR.

To begin the analysis, consider a photonic structure at temperature $T$, characterized by its spectral and angular emissivity $\varepsilon(\lambda, \theta)$. A theoretical heat transfer model, illustrated in Fig. S6, is developed to predict the stagnation temperature and net output energy of the TDPSR \cite{rephaeli2013ultrabroadband}. The structure is assumed to be exposed to a clear sky subjected to solar irradiance, as well as atmospheric irradiance corresponding to an ambient temperature $T_{amb}$. The net cooling power $P_{cooling}$ ($T$) of a structure with area $A$ is given by:

\begin{equation}
P_{cooling}\left(T\right)=P_{rad}\left(T\right)-P_{atm}\left(T_{amb}\right)-\ P_{sun}\ 
\end{equation}

The net heating power $P_{heating}$ ($T$) of the structure with area $A$ can be given by: 

\begin{equation}
P_{heating}\left(T\right)=P_{sun}+P_{atm}\left(T_{amb}\right)-P_{rad}\left(T\right)\ 
\end{equation}

where

\begin{equation}
P_{rad}\left(T\right)=A\bullet2\pi\int_{0}^{\frac{\pi}{2}}\sin{\theta}\cos{\theta}d\theta\int_{0}^{\infty}I_{BB}\left(T,\ \lambda\right)\ \varepsilon\left(\lambda,\ \theta\right)\ d\lambda
\end{equation}

is the power radiated by the structure, and

\begin{equation}
P_{atm}\left(T_{amb}\right)=\ A\bullet2\pi\int_{0}^{\frac{\pi}{2}}\sin{\theta}\cos{\theta}d\theta\ \int_{0}^{\infty}I_{BB}\left(T,\ \lambda\right)\ \varepsilon\left(\lambda,\ \theta\right)\ \varepsilon_{atm}\left(\lambda,\ \theta\right)\ d\lambda\ 
\end{equation}

is the power radiated by the atmosphere, which absorbs the power radiated by the structure

\begin{equation}
P_{sun}=A\int_{0}^{\infty}I_{AM1.5}\left(\lambda\right)\ \varepsilon\left(\lambda,\ 0\right)\ d\lambda
\end{equation}

is the solar power absorbed by the structure, the solar illumination is represented by $I_{AM1.5}$. We assume the structure is facing the sun. Therefore $P_{sun}$don’t need to have an angular integral, and the structure’s absorption is represented by its value in the zenith direction, so $\theta$ = 0.

\begin{equation}
I_{BB}\left(T,\ \lambda\right)=\ \frac{2hc^2}{\lambda^5}\bullet\frac{1}{e^\frac{hc}{\lambda k_bT}-1}
\end{equation}

is the spectral radiance density of a blackbody at temperature $T$.

\begin{equation}
\varepsilon_{atm}\left(\lambda,\ \theta\right)=1-\ {t(\lambda)}^\frac{1}{\cos{\theta}}
\end{equation}

is the angle-dependent emissivity of the atmosphere, where $t$($\lambda$) is the transmittance of the atmosphere in the zenith direction. We can replace the structure’s absorptivity with its emissivity $\varepsilon(\lambda, \theta)$ by using the Kirchoﬀ’s law. So Eq. 7 – Eq. 10 is universal for $P_{cooling}$ at high temperature and $P_{heating}$ at low temperature.

When the left sides of Eq. 5 and Eq. 6 are set to zero at steady-state, the equilibrium temperature Te can be determined by solving these Equations with the given environmental parameters (i.e., solar irradiation, ambient temperature, and atmospheric transmissivity) and the optical properties (i.e., absorptivity, reflectivity, and transmissivity) of TDPRS. Furthermore, the net cooling/ heating power of TDPRS can be calculated by solving Eq. 5 or Eq. 6 for the specified temperature setting of TDPRS.

\begin{figure}[h]
    \centering
    \label{tab:parameters}
    \includegraphics[width=0.8\textwidth]{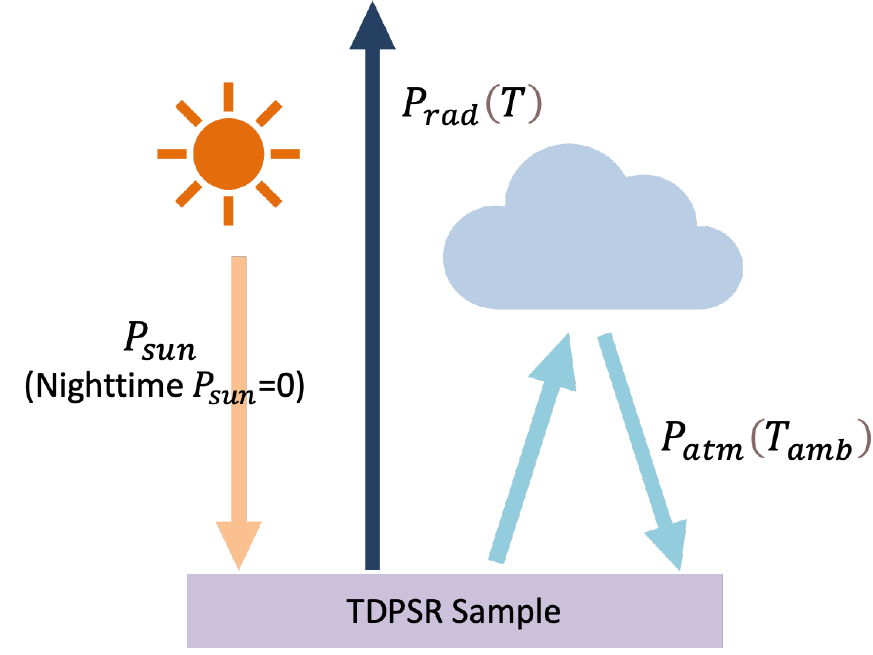} 
    \caption*{Fig. S6: Schematic diagram of thermal absorption and radiation}
\end{figure}

\section*{Acknowledgments}
First and foremost, I would like to express my deepest gratitude to my advisor, Jean-Jacque Delaunay, for giving me the opportunity to study at the University of Tokyo as an exchange student and to complete an independent project during my undergraduate studies. His unwavering support, invaluable guidance, and continuous encouragement throughout my research have been instrumental in the completion of this project. His insights and expertise have also played a crucial role in my success.

I would also like to extend my heartfelt thanks to Zhiyu Wang, the teaching assistant in the Delaunay lab. From providing the initial idea for this project, guiding me in finding relevant literature, discussing the feasibility of various options, to meticulously revising the manuscript. He played an indispensable role in helping me complete this paper. for his thoughtful feedback, constructive criticism, and for generously sharing his knowledge and expertise.

I am immensely grateful to my doctoral seniors Bo-Wei Lin, Di Xing, Mu-Hsin Chen, and Ying-Tsung Lee for their support and guidance with various simulation software. When I was struggling to use these tools for the first time, they willingly sacrificed their rest time to carefully guide me.

Lastly, I am grateful to all the participants and contributors to this research for their time and effort. Without their collaboration, this work would not have been possible.

\bibliographystyle{unsrt}  
\bibliography{ref}  

\end{document}